\documentclass[10pt,a4paper]{amsart}

\usepackage{enumerate}
\usepackage{amssymb}
\usepackage{exscale}
\newcommand{\be}{\begin{equation}}
\newcommand{\ee}{\end{equation}}
\newcommand{\bea}{\begin{eqnarray*}}
\newcommand{\eea}{\end{eqnarray*}}
\newcommand{\beq}{\begin{eqnarray}}
\newcommand{\eeq}{\end{eqnarray}}

\newcommand{\RR}{\mathbb{R}}
\newcommand{\NN}{\mathbb{N}}
\newcommand{\ZZ}{\mathbb{Z}}
\newcommand{\EE}{\mathbb{E}}
\newcommand{\PP}{\mathbb{P}}
\renewcommand{\r}{\right}
\renewcommand{\l}{\left}

\newcommand{\supp}{\mathop{\mathrm{supp}}}

\newcommand{\Tr}{\mathop{\mathrm{Tr}}}
\newcommand{\tL}{\mathop{\tilde\Lambda}}
\newcommand{\Lambdap}{\mathop{\Lambda^+}}

\newcommand{\la}{\langle}
\newcommand{\ra}{\rangle}
\newtheorem{thm}{Theorem}

\newtheorem{lem}[thm]{Lemma}
\newtheorem{cor}[thm]{Corollary}
\theoremstyle{remark}
\newtheorem{bem}[thm]{Remark}

\newenvironment{pro}{\begin{proof}}{\end{proof}}

\begin{document}

\title[Existence of the DOS for alloy-type potentials with small support]{Existence of the density of states\\ for one-dimensional alloy-type potentials\\ with small support}

\author{Werner Kirsch}
\author{Ivan Veseli\'{c}}
\address{Fakult\"at f\"ur Mathematik\\Ruhr-Universit\"at Bochum, Germany \hspace*{\fill}\linebreak
\mbox{and SFB 237  ''Unordnung und gro\ss e Fluktuationen''}}
\urladdr{http://www.ruhr-uni-bochum.de/mathphys/}
\dedicatory{Dedicated to the Memory of G.~A.~Mezincescu {\bf 1943---2001}} 

\date{\today}


\begin{abstract}
We study spectral properties of Schr\"odinger operators with random potentials of
alloy type on $L^2(\RR)$ and their restrictions to finite intervals. A Wegner estimate for non-negative single site potentials with small support is proven. It implies the existence and local uniform boundedness of the density of states. Our estimate is valid for all bounded energy intervals. Wegner estimates play a key role  in an existence proof of pure point spectrum.
\end{abstract}

\maketitle

\section{Model and results}

We study spectral properties of families of Schr\"odinger operators on  $L^2(\RR)$. The considered operators consist of a non-random periodic Schr\"odinger operator plus a random potential of {\it Anderson} or {\it alloy type}:
\beq
\label{H}
H_\omega := H_0 + V_\omega, \quad H_0 := -\Delta + V_{per}.
\eeq  
Here $\Delta $ is the Laplace operator on $\RR$ and $V_{per} \in L^\infty(\RR)$ is a $\ZZ$-periodic potential.  The random potential $V_\omega$ is a stochastic process of the following form
\begin{equation}
\label{V}
V_\omega (x) = \sum_{k \in \ZZ} \omega_k \,  u(x - k)
\end{equation}  
where $ \{ \omega_k \}_{k \in \ZZ} $ is a collection of independent identically
distributed random variables, called {\it coupling constants}. Their distribution
has a bounded density $f$ with support equal to a bounded interval. 
The non-negative {\it single site potential} $ u \in L^\infty(\RR)$ has compact support and an uniform positive lower bound on some open subset of $ \RR $. The potential $V:= V_{per} + V_\omega$ is (uniformly in $\omega$) bounded, hence $H_\omega$ is a selfadjoint, lower semibounded operator on the Sobolev space $W^{2,2}(\RR)$. For any interval $\Lambda_l:=\Lambda_l(x):= [-l/2,l/2]+x$ we can restrict $H_\omega$ to $L^2(\Lambda_l)$ with Dirichlet boundary conditions. We denote the restriction by $H_\omega^l$. It is again selfadjoint and lower semibounded and has discrete spectrum.

For $\mu(A) := \int_A f(x) dx $ define the product measure $\PP := \otimes_{k \in \ZZ} \, \mu$. We consider the collection $\{\omega_k\}_{k \in \ZZ}$ as an element of the probability space $(\Omega = \times_{k \in \ZZ}\RR, \PP)$. The expectation w.r.t.~$\PP$ is denoted by $\EE$.

By the general theory of ergodic random Schr\"odinger operators \cite{CarmonaL-90,PasturF-92} we know that there exists a $\omega$-independent set $\Sigma$ such that $\sigma(H_\omega)= \Sigma$ for $\PP$-almost all $\omega$. In the same way the spectral components $ \sigma_{ac},\sigma_{sc} $ and $ \sigma_{pp} $ are non-random subsets of the real line. Moreover, there exists a non-random distribution function $N$ called the {\em integrated density of states}  (IDS) which can be obtained by a macroscopic limit: Denote by
\begin{equation}
N_\omega^l (E) = l^{-d} \# \{ i | \ \lambda_i (H_\omega^l) < E \}
= l^{-d} \Tr P_\omega^l(]-\infty,E[) 
\end{equation}
the {\em finite volume IDS} or {\em normalized eigenvalue counting function} of $H_\omega^l$. Here $P_\omega^l(]-\infty,E[) $ denotes the spectral projection of $H_\omega^l$ on the energy interval $]-\infty,E[$. Then for all continuity points $E$ of $N$:
\begin{equation}
\label{ids}
N(E) = \lim_{l \to \infty} N_\omega^l (E) \quad \text{ $\PP$-almost surely}.
\end{equation}
Our main result reads:
\begin{thm}[Wegner estimate]
\label{theorem}
Let $H_\omega$ be as above. For any $E \in \RR$ there exist a constant $C$ such that 
\begin{equation}
\label{resultat}
\EE \l [ \Tr P_\omega^l( [E -\epsilon,E ]) \r ] 
\le C \, \epsilon \, l , \quad \forall \, \epsilon \ge 0.
\end{equation}
\end{thm}
Estimates of this type on the expectation value of the number of eigenvalues in a given energy interval go back to Wegner's paper \cite{Wegner-81} where he considers the discrete analog of $H_\omega$ on $l^2(\ZZ^d)$. 
Theorem \ref{theorem} proves the Lipschitz-continuity of the averaged finite volume IDS. So this function has a derivative at almost all energy values $E\in \RR$. 
Using \eqref{ids} this implies 
\begin{cor}
Under the assumptions of Theorem \ref{theorem} the IDS is Lipschitz continuous. Thus its derivative, the {\em density of states} $dN/dE$ exists for almost all $E\in \RR$ and it is uniformly bounded on any  interval $]-\infty, E]$.
\end{cor}
By the \v Ceby\v sev inequality (\ref{resultat}) implies
\begin{equation}
\label{Wegner}
\PP \l \{ \omega  \in  \Omega | \,  \sigma(H_\omega^l) \cap [E -\epsilon,E ] \neq \emptyset \r \}
\le C \epsilon \ l , \quad \forall \, \epsilon \ge 0.
\end{equation} 
For the application to the proof of localization only this form of the estimate is needed.
\begin{thm}[Localization]
\label{loctheorem}
Let $H_\omega$ be as in Theorem \ref{theorem} and $E\in\partial \sigma(H_\omega)$ any lower spectral edge. Then there exists an $\epsilon >0$ such that $H_\omega$ has $\PP$-almost surely pure point spectrum in $[E,E+\epsilon]$ with exponentially decaying eigenfunctions.
\end{thm}
We thus recover a result from \cite{KotaniS-87}, whose proof uses different methods.
In the present note we prove only Theorem \ref{theorem}, while the proof of Theorem \ref{loctheorem} can be 
found in the Diploma thesis \cite{Veselic-96}.
It uses the multiscale analysis of Fr\"ohlich and Spencer \cite{FroehlichS-83} in the version of von~Dreifus and Klein \cite{DreifusK-89}. Apart from the Wegner estimate \eqref{resultat}, a key ingredient is the Lifshitz asymptotics of the IDS derived by Mezincescu in \cite{Mezincescu-93} as well as a finite--infinite volume comparison lemma for the IDS \cite[Lem.~4.2]{Mezincescu-93}.
A detailed discussion of localization can be found e.g.~in \cite{Stollmann-2001} and the literature cited there. Section 3.4 of this monograph shows that Theorem \ref{loctheorem} can actually be strengthened to imply dynamical localization. For one-dimensional random operators there are several other techniques at disposal to prove localization, cf.~e.g~\cite{GoldsheidMP-77,CarmonaL-90,PasturF-92,BuschmannS-2001} and the references therein.
\begin{bem}
By assumption we have some open set $O$ and $\kappa>0$ such that $u \ge \kappa\chi_O$. For some $s>0$ the set $O\subset \RR$ contains a cube $\Lambda_s$ of sidelength $s$. By shifting the origin of $\RR$ we can therefore assume that $\Lambda_s$ has its center at $0$ and thus
 $
  u \ge \kappa \chi_{\Lambda_s(0)}.
 $
Moreover, by rescaling $f$ and $u$ we assume $\kappa =1$. 
Note that by adding a part of the periodic potential to $V_\omega$ we may assume without loss of generality that the support of $f$ starts at $0$, i.e. $ \supp f = [0, \omega_+] $ for some $ \omega_+ >0 $.
\end{bem}

Recently there has been increased interest in Wegner estimates for single site potentials $u$ that change sign \cite{Klopp-95a,Stolz-2000,Veselic-2000b+,HislopK-2001}. However even for nonnegative $u$ with small support the situation is not clearly understood. By ''small support'' we mean that 
\begin{equation}
\sum_{k \in \ZZ} u (x - k)
\end{equation}
is not bounded away from zero by a positive constant. Such potentials in arbitrary space dimension are considered e.g.~in \cite{Klopp-95a,Kirsch-96,BarbarouxCH-1997,KirschSS-1998a,CombesHN-2001}. However, the derived Wegner estimates are valid only at spectral boundaries.\footnote{See, however, the recent \cite{CombesHK-2002}.} 

Let us finish this section by mentioning that so far there is no Wegner estimate proven without the uniform positivity of $|u|$ on some open set. Namely, nothing is known in the case 
\begin{equation}
\label{notopen}
0 \le u \in L_c^2 (\RR),  \quad |\{ x | \, u(x) >0\} | > 0 
\end{equation}   
where $| \cdot|$ denotes Lebesgue measure. Property \eqref{notopen} implies the existence of some $\kappa >0$ such that $ U_\kappa :=\{ x | \, u(x) >\kappa\}  $ has positive measure. However $U_\kappa$ need not contain an open set. 

\section{Proof of Theorem \ref{theorem}\label{proof}}
Let us denote with $\rho : \RR \to [0,1]$ a smooth, monotone function with $ \rho = 1 $ on $ [ \epsilon ,  \infty [ $ and $ \rho = 0 $ on $ ] -\infty , -\epsilon ] $. The $n$-th eigenvalue of the operator $H_\omega^l$ is denoted by $E_n^l(\omega)$. We estimate similarly as in \cite[page 509]{Kirsch-96}
\begin{eqnarray*}
\EE \l [ \Tr P_\omega^l( [E -\epsilon,E+\epsilon ]) \r ]
\le
\int\dots\int  \prod_{k \in \Lambdap}   f(\omega_k) \: d\omega_k \sum_{n \in \NN} \int_{-2\epsilon}^{2\epsilon}d\theta 
\rho'(E_n^l(\omega) -E + \theta)  . 
\end{eqnarray*}
Here $\Lambdap := \{k \in \ZZ |  \, \supp \, u(\cdot-k) \text{ intersects } \Lambda_l  \}$ while  $\tL:= \ZZ \cap \Lambda_l$.
The above line can be bounded using the estimates from Section \ref{smallu} by 
\beq
\label{1par}
C_1 \int\dots\int  \prod_{k \in \Lambdap}   f(\omega_k) \: d\omega_k \sum_{n \in \NN} \int_{-2\epsilon}^{2\epsilon} d\theta 
 \sum_{ j \in \tL}
\frac{ \partial \rho (E_n^l(\omega) -E + \theta) }{ \partial \omega_j } 
\\
\nonumber
=C_1  \sum_{ j \in \tL} \int\dots\int  \prod_{k \in \Lambdap\setminus j}   f(\omega_k) \: d\omega_k \sum_{n \in \NN} \int_{-2\epsilon}^{2\epsilon}  d\theta \int  d\omega_j \,  f(\omega_j)
\frac{ \partial \rho (E_n^l(\omega) -E + \theta) }{ \partial \omega_j }  . 
\eeq
using  Beppo Levi's theorem. Denote by $H^l(j, \omega_+) $ the operator $H_\omega^l$ where the random variable $\omega_j$ has been set to its maximum value $\omega_+$, and similarly $H^l(j, 0) $.  The eigenvalues of the two above operators are abbreviated by $E_n^l(j,\omega_+)$, resp.~$E_n^l(j,0)$. 
Using monotonicity we estimate from above the sum over $n$ in the second line of (\ref{1par}) by
\begin{equation}
\label{TrEst}
\|f\|_\infty \sum_{n \in \NN} \int_{-2\epsilon}^{2\epsilon}  d\theta 
\l \{ \rho (E_n^l(j,\omega_+) -E + \theta)- \rho (E_n^l(j,0) -E + \theta)\r \}
\end{equation}
Let the single site potential $u$ be supported in $[-R,R]$.
Introduce now the operators $H_\omega^{l,*}$, $*\!=\!D,N$  which coincide with $H_\omega^{l}$  up to additional Dirichlet, respectively Neumann b.c.~at the points $j\!-\!R$ and $j\!+\!R$. Their eigenvalues are $E_n^{l,*}(j,\omega_+)$. By Dirichlet-Neumann bracketing, the braces in \eqref{TrEst} are bounded by 
\beq
\label{DNb}
\rho (E_n^{l,D}(j,\omega_+) -E + \theta)- \rho (E_n^{l,N}(j,0) -E + \theta).
\eeq
As $H_\omega^{l,*}= H_\omega^{j,*} \bigoplus H_\omega^{c,*}$ is a direct sum of two operators acting on $L^2(j\!-\!R,j\!+\!R)$ and $L^2(\Lambda_l \setminus [j\!-\!R,j\!+\!R])$ the sum over the terms in \eqref{DNb} can be separated:
\beq
\label{c}
&&\sum_n \rho (E_n^{c,D}(j,\omega_+) -E + \theta)- \rho (E_n^{c,N}(j,0) -E + \theta).
\\
\label{j}
&+&
\sum_n \rho (E_n^{j,D}(j,\omega_+) -E + \theta)- \rho (E_n^{j,N}(j,0) -E + \theta).
\eeq 
Since the the difference in the boundary conditions is a rank two perturbation in resolvent sense (see e.g.~\cite{Simon-95f}), the interlacing theorem says that the first term in  \eqref{c} is bounded by $\rho (E_{n+2}^{c,N}(j,\omega_+) -E + \theta)$. A telescoping argument bounds the whole sum in \eqref{c} by twice the total variation of $\rho$, which is equal to one. The sum in \eqref{j} we estimate by 
\begin{multline*}
\Tr\l [\chi_{[E-3\epsilon ,\infty[} (H_\omega^{j,D}(j, \omega_+))- \chi_{]E+3\epsilon,\infty[} (H_\omega^{j,N}(j,0))\r ] 
\\
\le 
2 + 
\Tr\l [\chi_{[E-3\epsilon ,\infty[} (H_\omega^{j,D}(j,0)+\|u_j\|_\infty)- \chi_{]E+3\epsilon,\infty[} (H_\omega^{j,D}(j,0))\r ]
\le
C_2,
\end{multline*}
where the constant $C_2$ depends only on $E+3\epsilon, V_{per}, \omega_+$ and $u$, cf.~\cite{KirschM-82c}.
The proof of Theorem \ref{theorem} is finished by the the upper bound on  (\ref{1par}):
\begin{eqnarray*}
C_1 \|f\|_\infty \sum_{ j \in \tL} \int\dots\int  \prod_{k \in \Lambdap\setminus j}   f(\omega_k) \: d\omega_k 
\, (C_2+2) \,4\epsilon \, l
= 4 C_1 (C_2+2) \|f\|_\infty \, \epsilon  \, l.
\end{eqnarray*}

\section{Single site potentials of small support}
\label{smallu}
In this section we prove the uniform lower bound 
\begin{equation}
\label{Eder}
\sum_{k \in \tL} \frac{\partial E_n^l(\omega)}{\partial \omega_k} \ge C_4(I) >0
\end{equation}
for all eigenvalues $E_n^l$ of $H_\omega^l$ inside a bounded energy interval $I$. The bound $C(I)$ does not depend on the sidelength $l \in\NN$ and on the eigenvalue index $n\in \NN$.
By the chain rule 
\begin{eqnarray*}
 \sum_{ j \in \tL_l}
\frac{ \partial \rho (E_n^l(\omega) -E + \theta) }{ \partial \omega_k } 
=
\rho'(E_n^l(\omega) -E + \theta)  
\sum_{k \in \tL_l} \frac{\partial E_n^l(\omega)}{\partial \omega_k}
\end{eqnarray*}
\eqref{Eder} implies the estimate needed in Section \ref{proof}:
\begin{eqnarray*}
\rho'(E_n^l(\omega) -E + \theta) 
\le
C_4(I)^{-1}
 \sum_{ j \in \tL}
\frac{ \partial \rho (E_n^l(\omega) -E + \theta) }{ \partial \omega_j } .
\end{eqnarray*}
To infer the lower bound \eqref{Eder} set $S= \bigcup_{k\in \tL} (\Lambda_s(k))$ and apply the Hellman-Feynman theorem. For a normalized eigenfunction $\psi_n$ corresponding to $E_n^l(\omega)$:
\begin{eqnarray*}
\sum_{k \in \tL} \frac{\partial E_n^l(\omega)}{\partial \omega_k}
=
\sum_{k \in \tL} \la \psi_n, u(\cdot-k)\psi_n \ra
\ge 
\int_{S} |\psi_n|^2.
\end{eqnarray*} 
If the integral on the rightern side would extend over the whole of $\Lambda_l$ it would be equal to $1$ due to the normalization of $\psi_n$.  A priori the integral over $S$ could be much smaller, but the following Lemma shows that this is not the case. 
\begin{lem}
\label{uc}
Let $I$ be a bounded interval and $s>0$. There exists a constant $c>0$ such that 
\begin{eqnarray*}
\int_{\Lambda_s(k)} |\psi|^2 \ge  c\int_{\Lambda_1(k)} |\psi|^2
\end{eqnarray*} 
for  all $l\in \NN$, all $k\in \tilde\Lambda_l$ and for any eigenfunction $\psi$ corresponding to an eigenvalue $E\in I$ of $H_\omega^l$.
\end{lem}
Thus $\int_{S} |\psi|^2 \ge  c\int_{\Lambda_l} |\psi|^2$ with the same constant as in Lemma \ref{uc}.
\begin{pro}
For 
\begin{equation}
 \phi ( y) := \int_{\Lambda_s(k+y)} dx \, |\psi(x)|^2  = \int_{\Lambda_s(k)} dx \, |\psi(x-y)|^2 
\end{equation}
 one has
\begin{eqnarray*}
\l | \frac{\partial}{\partial y}  \phi ( y)   \r | 
& = & \l |  \int_{\Lambda_s(k)} dx \, \l [  \frac{\partial}{\partial y} \psi (x-y)  \r ] \overline{ \psi(x-y) } 
        + \int_{\Lambda_s(k)}  dx  \, \psi(x-y) \, \frac{\partial}{\partial y} \overline{ \psi(x-y) } \r | 
\\
& \leq & 2 \l \|  \psi \r \|_{ L^2( \Lambda_s(k+y) ) } \l \| \psi'\r \|_{ L^2( \Lambda_s(k+y) ) }.
\end{eqnarray*}
Sobolev norm estimates (e.g.~Theorems 7.25 and 7.27 in \cite{GilbargT-83}) imply
\[
\|  \psi' \|_{L^2( \Lambda_s(k+y) )} \le  C_5 \,  \|  \psi \|_{L^2( \Lambda_s(k+y) )} +\|  \psi'' \|_{L^2( \Lambda_s(k+y) )}.
\]
By the eigenvalue equation  we have
\begin{equation}
\l | \frac{\partial}{\partial y}  \phi ( y)   \r | 
\le C_6 \  \|  \psi  \|_{ L^2( \Lambda_s(k+y) ) }^2 = C_6 \, \phi ( y), \qquad C_6 =C_6(\|V-E\|_\infty).
\end{equation}
Gronwall's Lemma  implies  $ \phi(y)  \le  \exp(C_6 |y| ) \, \phi(0)$ and thus 
\begin{eqnarray*}
\int_{\Lambda_1(k)} |\psi|^2 \le  e^{C_6}    \ s^{-1} \int_{\Lambda_s(k)} |\psi|^2. 
\end{eqnarray*}
\end{pro}
At the Conference in Taxco we learned from J.M.~Combes that there is a article in preparation together with P.~Hislop and F.~Klopp with slightly weaker results as in this work applying also to the higher dimensional case \cite{CombesHK-2002}.

\def\cprime{$'$} \def\cprime{$'$}
\providecommand{\bysame}{\leavevmode\hbox to3em{\hrulefill}\thinspace}
\providecommand{\MR}{\relax\ifhmode\unskip\space\fi MR }
\providecommand{\MRhref}[2]{%
  \href{http://www.ams.org/mathscinet-getitem?mr=#1}{#2}
}
\providecommand{\href}[2]{#2}

\end{document}